\documentclass[superscriptaddress, reprint, amsmath, amssymb, aps, pra, floatfix]{revtex4-2}

\usepackage[colorlinks=true]{hyperref}
\usepackage{graphicx}
\usepackage{dcolumn}
\usepackage{bm}
\usepackage{orcidlink}
\usepackage{physics}
\usepackage{color}
\usepackage{amsthm}

\begin{document}


\title{Quantum education in the frontier city of Kharkiv}

\author{{Pylyp} {Kuznietsov}\orcidlink{0000-0001-8477-1395}}\email{kuznietsov@karazin.ua}
\affiliation{V.N. Karazin Kharkiv National University, Svobody Square 4, 61022 Kharkiv, Ukraine}

\author{{Igor} {Girka}\orcidlink{0000-0001-6662-8683}}
\affiliation{V.N. Karazin Kharkiv National University, Svobody Square 4, 61022 Kharkiv, Ukraine}

\author{{Igor} {Kyryllin}\orcidlink{0000-0003-3625-7521}}
\affiliation{Akhiezer Institute for Theoretical Physics, NSC KIPT, Akademichna 1, 61108 Kharkiv, Ukraine}
\affiliation{V.N. Karazin Kharkiv National University, Svobody Square 4, 61022 Kharkiv, Ukraine}

\author{Andrii Sotnikov\orcidlink{0000-0002-3632-4790}}
\email{a\_sotnikov@kipt.kharkov.ua}
\affiliation{Akhiezer Institute for Theoretical Physics, NSC KIPT, Akademichna 1, 61108 Kharkiv, Ukraine}
\affiliation{V.N. Karazin Kharkiv National University, Svobody Square 4, 61022 Kharkiv, Ukraine}

\date{\today}

\begin{abstract}
This article provides the description of current training processes and structure of education in quantum physics at the Education and Research Institute ``School of Physics and Technology'' (SPT) of V.N.~Karazin Kharkiv National University, Ukraine. Crucial feature of quantum education at the SPT is the  involvement of scientists and experts from national and international research centers who are actively working in the field. By taking example of particular quantum courses, we outline the main challenges in the educational process during the large-scale military aggression and the ways the lecturers, scientific employees, and students manage to overcome them. We also overview the recently emerged initiatives oriented on sustaining and development of quantum education in Kharkiv, as well as the international events with a broader impact. History of the School and its main achievements are provided in brief.
\end{abstract}

\maketitle 

\section{Introduction} 

Metropolitan city of Kharkiv, with its population of about 1.5 million, is a major urban hub boasting a substantial pool of scientists, engineers, highly-skilled IT individuals, and a vibrant student community. It is a host city for a number of research institutions and universities with many researchers actively working in the field of quantum science and technology.
Our aim is not only to outline the historical steps in formation of the quantum education and science in the city together with the faced challenges since the full-scale military aggression in 2022, but to rejuvenate and inspire the physics community in Kharkiv, heavily affected by direct confrontation during the war in Ukraine. It is committed to uplifting and motivating the local research associates, contributing to the resurgence of the scientific mind in the region, as well as to bring supportive ideas and potential joint programs for revitalization from the worlds' scientific community.

\subsection{Brief history of quantum education at the School of Physics and Technology, Kharkiv University}\label{subsec.1.1}

In 1925--1928, the communist regime carried out another flagrant social experiment in USSR. All Universities were closed in Ukraine. It was assumed that classical universities are to be substituted by institutes concentrating on specific purposes of the authorities. The idea was as follows. A person is a small ``cog'' in a giant machine being the state. According to the paradigm, a regular professor should study at the University, publish several research papers, defend a Ph.D. thesis, and deliver lectures, e.g., on Quantum Mechanics till retiring. 

In 1943, a large-scale program on the creation of atomic weapons began in the USSR. It appeared that the ``scientists'', educated according the model described above, were disabled to contribute to the problem solving. In spite of this, Kharkiv Institute of Physics and Technology (KIPT) has accepted the challenge and was among the research centers, which took part in the Soviet atomic project. 

On January 28, 1946, the Decree of the Soviet government No. 225-96cc (cc stands here for the ``top secret'') ``On the training of physical engineers and specialists in atomic nuclear physics and on radiochemistry'' was signed by I. Stalin. The Decree has assigned duties on specialists training to Moscow, Leningrad, Kyiv and Kharkiv state universities, as well as to the Moscow mechanical and Leningrad polytechnic institutes. This was the starting point of emerging new field of science and technology (i.e., the nuclear physics) in Kharkiv. 

The education at Kharkiv University was arranged at the Department of Electronic Processes and Nuclear Physics (since 1946), then Special (nuclear) Department of the School of Physics and Mathematics (since 1948). In 1962, the Department was rearranged into the School of Physics and Technology (SPT). 

\subsection{The essence of education at SPT}\label{subsec.1.2}

The nuclear students were taught by professionals -- those people from KIPT, who did science themselves. This was and remains the crucial point in arranging the education at SPT by now. Moreover, the students of SPT are involved into carrying out research projects at KIPT and other physical institutes of National Academy of Science of Ukraine (NASU) already from the third/fourth year of studies. In other words, BSc and MSc theses of the students are mostly in course of advanced Physics investigated at KIPT. Modern physical equipment and instruments, available at KIPT, are used for developing the practical competences of the students. At the moment, more than 25\% of professors at SPT are part-time employed professors, whose main employment corresponds to the research positions at KIPT. Recently the University has accepted the concept of visiting lecturers that gave the opportunity for the scientists and professors from the research centers and institutions from abroad to officially lecture for SPT students. We mark out that those visiting professors from abroad institutions are mostly SPT graduates who became a leading experts at foreign research centers. This concept can be concidered as the extension of classical SPT education essence. 

Three Departments of SPT are headed by the members of NASU. This means that the lecturers provide students not only with the contents of the textbooks or other published materials, but also the most recent results in their field of expertise and everyday research activity.
The curriculum at SPT is arranged and being improved permanently according to the modern trends in physical research.

\subsection{The achievements of SPT graduates and students}\label{subsec.1.3}

The outlined approach (joint efforts of the research centers and the University) to the arrangement of the training demonstrates impressive results. 
In particular, approximately one third of SPT graduates obtains PhD degrees after graduation. Thirty-three SPT graduates were elected to the NASU, and academies of other countries, which is more than a half of those elected from the University as a whole. Since 1991, eighteen SPT graduates were awarded the State Prize of Ukraine in the field of science and technology. 

Students of SPT have won the All-Ukrainian Olympiads in Physics more than twenty times. In addition, they won eight times the All-Ukrainian Olympiads in the related specialties: theoretical mechanics, astronomy, astrophysics, and the English language. In the period from 2003 to 2019, eighteen All-Ukrainian student physics tournaments were held. SPT students won prizes in all the tournaments, including ten times when they won first degree diplomas. Since 2011, when the International Physicists’ Tournament (IPT) ceased to be held in Ukraine, until 2020, SPT students have represented Ukraine at the IPT six times~\cite{IntPhysTourn}. And every time they returned home with a victory, and three times they brought home gold medals.

SPT undergraduates won sixteen times in the final rounds of the All-Ukrainian competition for student scientific papers in the fields of ``physics'', ``technology of construction materials and materials science'', and ``mechanical engineering''. During 2016-2020, SPT students co-authored 79 papers published in journals indexed by the Scopus. That is, there were 0.57 papers for each SPT undergraduate.

\section{Education during the war}\label{sec2}

\subsection{Consequences of the full-scale Russian aggression against Ukraine for SPT}\label{subsec.1.3}

\begin{figure*}[t]
\centering
\includegraphics[width=\textwidth]{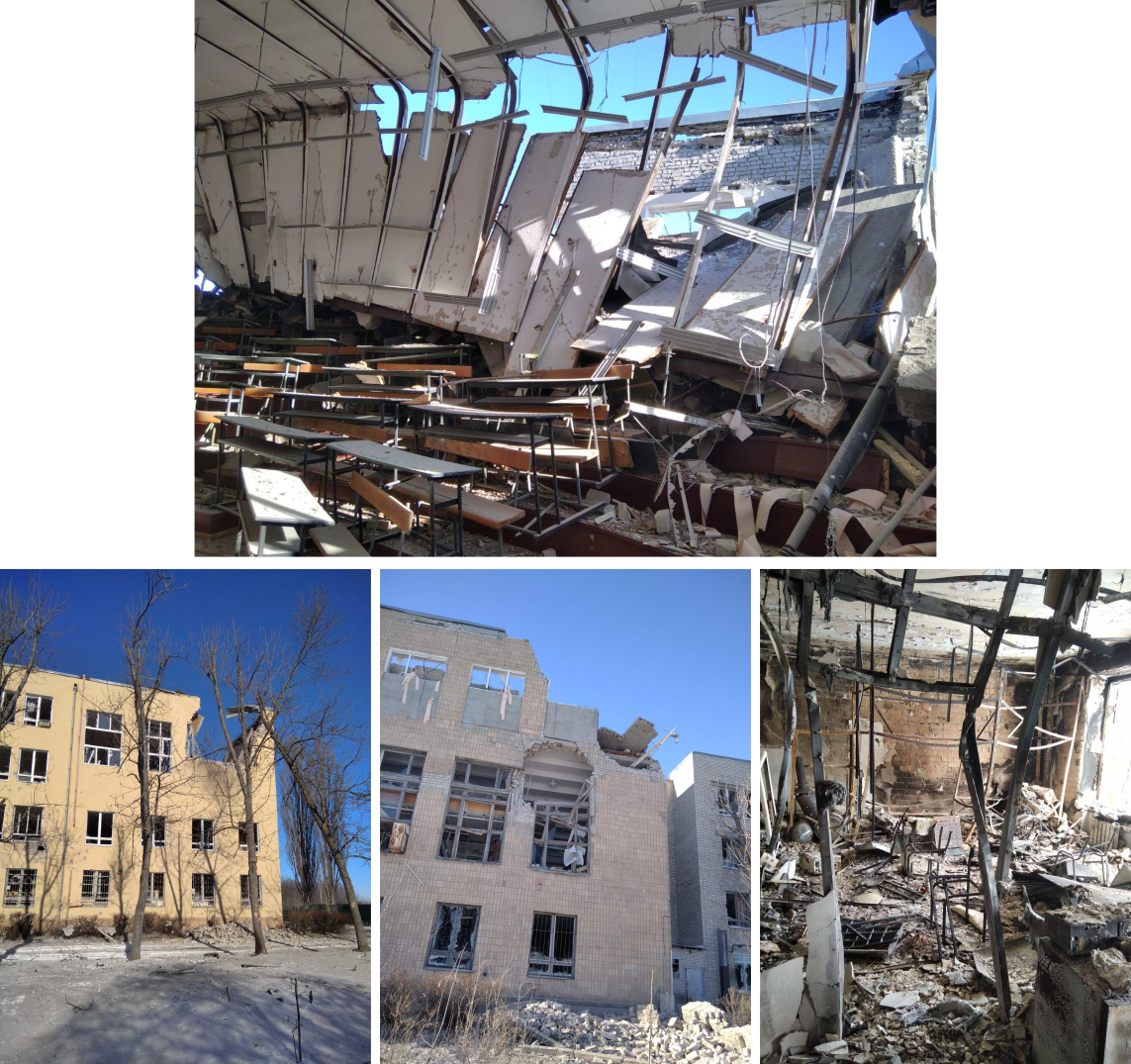}
\caption{The state of the SPT educational building as of March 2024: the lecture hall, the view from East, the view from North, and the laboratory of molecular spectroscopy.
}\label{fig1}
\end{figure*}

The SPT campus is located separately from the main building of the Karazin University. It stands on the Northern outskirt of Kharkiv city, in other words, it is one of the closest city structures to the Ukrainian state boarder. The building was partly destroyed by Russian missiles already during the first night of the Russian invasion, on February 24-25, 2022. Not only research labs and lecture halls were destroyed, as shown in Fig.~\ref{fig1}. Student dormitory was hit by Russian missiles as well. Some of SPT professors, engineers and students voluntary joined Ukrainian Armed Forces to defend the Motherland. About 60\% of teaching staff of Karazin University had to move from Kharkiv to Western cities of Ukraine or abroad. But those who stayed in the city had to overcome this terrifying circumstances. 

At the beginning of the invasion, the main task was to preserve the scientific, educational and research infrastructure. Significant efforts were undertaken to relocate the extant equipment of the SPT to the safe locations at KIPT and other University buildings. This is another example of positive interaction between the University and the Institute. According to the safety requirements the education process at the University is carried out on-line only. This led to the necessity to provide researchers, lecturers, and tutors with the digital learning tools and to teach them new learning methods. Nevertheless, it is obviously unsatisfactory for students-physicists to study without practical exercises on real physical installations, without measuring physical observables themselves. This issue is resolved by attending them laboratories in KIPT and partner institutions abroad.

Despite the partial destruction of the SPT building, some of the labs are still operating there. The administration of the University has already prepared the reconstruction project of the building. It will cost approximately 8 million euro not accounting costs to purchase research equipment. Total damages to Karazin University are estimated at more than 150 million euro.

\section{Quantum courses at SPT}\label{sec3}
The studying of quantum physics at SPT begins in the second (spring) semester of the third year when students start studying Quantum Mechanics. An important element of the SPT education system is that most disciplines are taught by former SPT graduates. For over 50 years, quantum mechanics at SPT has been taught by Prof. Yuri Berezhnoy, who was a student of the renowned Ukrainian physicist, Academician Oleksandr Akhiezer.
It is in place to remind that Oleksandr Akhiezer was one of Lev Landau's closest collaborators and an active member of his world-famous scientific school.

In 1953, Yuri Berezhnoy entered the School of Physics and Mathematics at Kharkiv University. After the first two years of study, he was admitted to the Special (nuclear) Department. Quantum Mechanics was taught to him by O.I. Akhiezer (and before him, this course was taught at the School of Physics and Mathematics by L.M. Pyatigorsky). After graduating from Kharkiv University, Yuri Berezhnoy worked at KIPT. And in 1979, he became the head of the Department of Theoretical Nuclear Physics at SPT, teaching Quantum Mechanics until 2021. Currently, Quantum Mechanics is also taught by SPT graduate DSc Igor Kyryllin.

The study of Quantum Mechanics at SPT spans two semesters. In the first semester, students attend two lectures and work on one practical session per week. The first semester begins with the studying of the physical and mathematical foundations of quantum mechanics. Students become familiar with the fundamental ideas and principles of Quantum Mechanics: the concept of quantization, the corpuscular-wave duality of microobjects, the uncertainty principle, and the principle of superposition. They also learn about the wave function and its physical meaning, studying the superposition of states and the wave function of non-interacting microobjects. The theoretical knowledge gained in lectures is reinforced through practical sessions, where students solve problems independently or with the help of the professor.

Subsequently, students delve into the mathematical foundations of Quantum Mechanics, covering topics such as operators of physical quantities, continuous and discrete spectra, the formalism of state vectors, the Hamiltonian operator, the description of the evolution of a quantum system in time, and working with the density matrix. The course progresses to the study of the momentum operator, its eigenfunctions and commutation relations, uncertainty relation, and the concept of a wave packet. The notion of the angular momentum operator is introduced, including its eigenvalues and eigenfunctions, addition rules for angular momenta, the concept of parity, spin, and wave functions of particles with spins.

Following this, students study the Schr{\"o}dinger equation and explore topics such as one-dimensional motion, potential barrier, potential well, oscillator (in coordinate and number representation), coherent states of the oscillator, and three-dimensional oscillator. The course then covers the motion of a particle in an external field, two-body problem, free motion, finite motion of a charged particle in an electric field, motion of a charged particle in a magnetic field, and Landau levels.

At the end of the first semester, students delve into systems of identical particles, the indistinguishability of identical particles, wave functions with specific symmetry, exchange interaction, and second quantization for systems of identical bosons and fermions. Approximation methods such as perturbation theory and semiclassical approximation are also introduced.

In the second semester, students attend one lecture and one practical session per week. They study the electronic structure of atoms, self-consistent field, Thomas-Fermi equation, periodic table of elements, multipole moments, atoms in electric and magnetic fields, molecular structure, valence, and Van der Waals forces.

Subsequently, students explore the quantum scattering problem, including partial wave method, scattering matrix, slow particle scattering, scattering of slow neutrons by protons, inelastic collisions, Born approximation, high-energy approximation, scattering of charged particles in an electric field, scattering of high-energy electrons by atoms, scattering of high-energy electrons by atomic nuclei, scattering of identical particles, and spin effects in scattering.

In addition to practical sessions, students are required to solve quantum mechanics problems during self-preparation. Each student presents the solutions to these problems to the professor, utilizing an online whiteboard during seminar sessions.

After studying Quantum Mechanics, SPT students continue with many other quantum disciplines, including Quantum Electrodynamics, Statistical Physics, Elementary Particle Physics, Nuclear Physics, Quantum Field Theory, Physics of the Early Universe, High-Energy Physics and Standard Model, Electromagnetic Processes in High-Energy Physics, Quantum Many-body Systems and Critical Phenomena, Quantum Information and Mesoscopic Physics, Solid State Theory, etc.

Unfortunately, the war complicated the educational process. Of course, working with students in the classroom is much more interesting and efficient than in front of a computer screen. But we try to teach the material as efficiently as possible despite the circumstances. Zoom and online boards greatly assist in this respect. Nevertheless, we hope that in the near future, we will be able to return to lecture halls and work with real, not virtual, boards.

It is worth noting that there are modern textbooks and monographs for several courses taught at SPT, which are authored by the School's professors. For instance, the foundational textbook on Quantum Mechanics \cite{Berezhnoy2014}, as well as monographs related to the courses on Quantum Information Theory \cite{Shevchenko2019}, Quantum Statistical Mechanics~\cite{Sotnikov2023}, Quantum Scattering Theory~\cite{Berezhnoy2013} and Theory of Nuclear Reactions~\cite{Berezhnoy2011}.

\section{Recently emerged initiatives and the future}\label{sec4}

In 2023, following the calls of the time, several key initiatives appeared, which are aimed at intensifying quantum education and research at SPT.
Quantum technologies play ever increasing role in the modern economy and, thus, these initiatives may not only develop and further strengthen the existing international collaboration, but also lay the ground for Ukraine’s revitalization.

First, in May 2023 we contributed to the formation of a quantum workgoup (qUA) of the Ukrainian Physical Society.
This group appeared on a voluntary basis and is formed by five representatives of Kharkiv, Kyiv, Lviv, and Donetsk \cite{qUA}. Initially, the aim was to promote Ukrainian scientists working in the field of quantum physics to become more visible on the EU quantum research landscape, in particular, for the Quantum Flagship program (EU Horizon Europe Programme) \cite{qFlagship}, so that they could contribute on equal basis either to the existing research programs (via the hop-on opportunity) or participate in the future calls for the Quantum Flagship projects. However, soon we realized the necessity to extend contribution of qUA in order to inform the focus community in Ukraine about the research and educational initiatives, seminars, calls for projects, papers, open research positions, schools, and conferences.

As the second milestone, SPT contributed to the organization of the US-Ukraine Quantum Forum, August 28-31, 2023 \cite{bib3}. The purpose of this event was to bring together quantum researchers from the United States and Ukraine. In total, 32 speakers presented their talks during four days; the sessions were actively attended by more than 50 Ukrainian and US scientists and students. This event was especially inspiring for the Ukrainian researchers falling under travel restrictions for the male category. The speakers residing in Ukraine were additionally supported in the form of honorarium by the Office of Naval Research Global Grant \# N62909-23-1-2088  (program manager Martina Barnas).

After resounding success of the quantum forum, we realized the necessity to keep high level of scientific exchange with foreign colleagues. By joining the efforts from three Kharkiv institutions, we organized the Kharkiv Quantum Seminar series, which are held on a regular basis in the virtual format since October 2023 \cite{KQS}. The seminars are given dominantly by the speakers on the middle stages of their research careers, to a large extent, by the exceptionally motivated representatives of Ukrainian scientific diaspora (Denys Bondar, Sergey Frolov, Kyrylo Snizhko, Artem Volosniev, Olexandr Isayev, Denis Seletskiy, {\it et al.}). At the same time, there are also invited speakers, the world-leading specialists in quantum physics, such as Steven M. Anlage, Tomasz Dietl,  Gediminas Juzeliūnas, Michael Berry, Jan Kune\v{s}, John Perdew, Franco Nori, Roger Melko, {\it et al.}, who expressed their sincere interest in giving lectures in these seminars. In addition, there are weekly seminars at SPT, with speakers being mostly researchers on early stages of their scientific careers (dominantly formed by recent graduates from SPT) sharing their knowledge with colleagues and researchers in Kharkiv.

SPT has recently introduced new BSc educational program on ``Cyberphysical nuclear technologies''. It contains classical quantum courses as well as newly introduced ones, such as ``Quantum computer basics'' and ``Quantum algorithms''. Currently, we are on the stage of development of the Master-level educational program oriented on quantum computing and quantum technologies. In particular, by joint efforts of representatives of Taras Shevchenko Kyiv National University (Maksym Teslyk) and University of South Carolina (Yaroslaw Bazaliy) the SPT students are attending the new course on quantum information theory. Also, SPT students participated in the winter school on quantum computing ``QHack UA Bootcamp 2024'' \cite{QHack2024UA}, where their team won the first prize. After that, they participated as the ``Kharkiv\_team'' in the international QHack 2024 event, where they won the first prize in the challenge ``A Matter of Taste'' and were among top-5 teams in the open hackaton competition \cite{QHack2024}.

As a bitter pill for the quantum community in Ukraine and, especially, in the frontier city of Kharkiv, we were heavily affected to learn about the specific details of the resolution for the 2023 General Conference of the United Nations Educational, Scientific and Cultural Organization (UNESCO) and the 2023 General Assembly of the United Nations to proclaim 2025 the International Year of Quantum Science and Technology, in particular, under the tab ``our partners'' \cite{QYear}. As of March 2024, the corresponding webpage indicates that their partners are, in particular, the ``Russian Academy of Sciences'' (RAS) and ``Russian Physical Society'' (RPS).
Also, the academic community should be aware of the fact that about 200 rectors (presidents) of Russian universities, led by Professor Viktor Sadovnichy, Rector of Moscow State University, signed a letter that justifies, approves and encourages war in Ukraine \cite{Rectors}.

As outlined above, having witnessed the irrevocable impact of the military aggression on education and science in Kharkiv, the authors express deep concerns about the existence of outweighing positive contributions of the listed two organizations to the development or, at least, to sustaining Quantum Science and Technology on the global scale since February 24, 2022.
To be more precise, on the RAS and RPS websites there is no information about their official position on the war (called there as a ``special military operation'') and massive destruction of the research and educational infrastructure in Ukraine and Kharkiv, in particular. The absence of any special opinion on this matter seems to us to be full and categorical support of the state in this situation. A valid question arises to the partners: Is that an appropriate company to celebrate the International Year of Quantum Science and Technology with?

\section{Conclusion}

Quantum education at the School of Physics and Technology, V.N. Karazin Kharkiv National University, Ukraine, has a long history. Its foundations were laid by Academicians Lev Landau and Oleksandr Akhiezer. Quantum education was and is provided here by the researchers who are actively working in modern Quantum Physics. The efficiency of this distinction is proved by successful scientific trajectories of the School’s graduates. Full-scale Russian aggression destroyed the buildings of the School and students’ dormitory and seriously complicated the training, especially its practical components. However, the aggression did not break our spirit and determination to defend our freedom and independence. For more than two years, Ukraine, with the support of the democratic world, has resisted full-scale Russian aggression, and Ukrainian scientists continue demonstrating top-level scientific results.

%
%

\begin{acknowledgments}

The authors deeply appreciate the Ukrainian Armed Forces for the possibility to continue teaching students and carrying out research. Comprehensive support of the democratic world to Ukraine in its struggle against Russian invasion is greatly acknowledged.

\end{acknowledgments}


%

\end{document}